\documentclass[acmsmall]{acmart}

\AtBeginDocument{%
  \providecommand\BibTeX{{%
    \normalfont B\kern-0.5em{\scshape i\kern-0.25em b}\kern-0.8em\TeX}}}

\setcopyright{rightsretained}
\acmJournal{PACMHCI}
\acmYear{2023} \acmVolume{7} \acmNumber{CSCW1} \acmArticle{136} \acmMonth{4} \acmPrice{}\acmDOI{10.1145/3579612}

\received{July 2022}
\received[revised]{October 2022}
\received[accepted]{January 2023}

\usepackage{layouts}
\usepackage{enumitem}
\usepackage{amsmath}
\usepackage[noabbrev,capitalize,nameinlink]{cleveref}
\usepackage{ulem}

\newcommand{\alex}[1]{}
\newcommand{\adam}[1]{}
\newcommand{\jason}[1]{}

\begin{document}

\title{Improving Human-AI Collaboration with Descriptions of AI Behavior}


\author{Ángel Alexander Cabrera}
\email{cabrera@cmu.edu}
\orcid{0000-0003-0348-3362}
\affiliation{%
  \institution{Carnegie Mellon University}
  \streetaddress{5000 Forbes Ave}
  \city{Pittsburgh}
  \state{Pennsylvania}
  \country{USA}
  \postcode{15213}
}

\author{Adam Perer}
\email{adamperer@cmu.edu}
\orcid{0000-0002-8369-3847}
\affiliation{%
  \institution{Carnegie Mellon University}
  \streetaddress{5000 Forbes Ave}
  \city{Pittsburgh}
  \state{Pennsylvania}
  \country{USA}
  \postcode{15213}
}

\author{Jason I. Hong}
\email{jasonh@cs.cmu.edu}
\orcid{0000-0002-9856-9654}
\affiliation{%
  \institution{Carnegie Mellon University}
  \streetaddress{5000 Forbes Ave}
  \city{Pittsburgh}
  \state{Pennsylvania}
  \country{USA}
  \postcode{15213}
}

\renewcommand{\shortauthors}{Cabrera et al.}

\begin{abstract}
People work with AI systems to improve their decision making, but often under- or over-rely on AI predictions and perform worse than they would have unassisted.
To help people appropriately rely on AI aids, we propose showing them \textit{behavior descriptions}, details of how AI systems perform on subgroups of instances.
We tested the efficacy of behavior descriptions through user studies with 225 participants in three distinct domains: fake review detection, satellite image classification, and bird classification.
We found that behavior descriptions can increase human-AI accuracy through two mechanisms: helping people identify AI failures and increasing people's reliance on the AI when it is more accurate.
These findings highlight the importance of people's mental models in human-AI collaboration and show that informing people of high-level AI behaviors can significantly improve AI-assisted decision making.
\end{abstract}




\maketitle

\section{Introduction}

\begin{figure*}
    \centering
    \includegraphics[width=\textwidth]{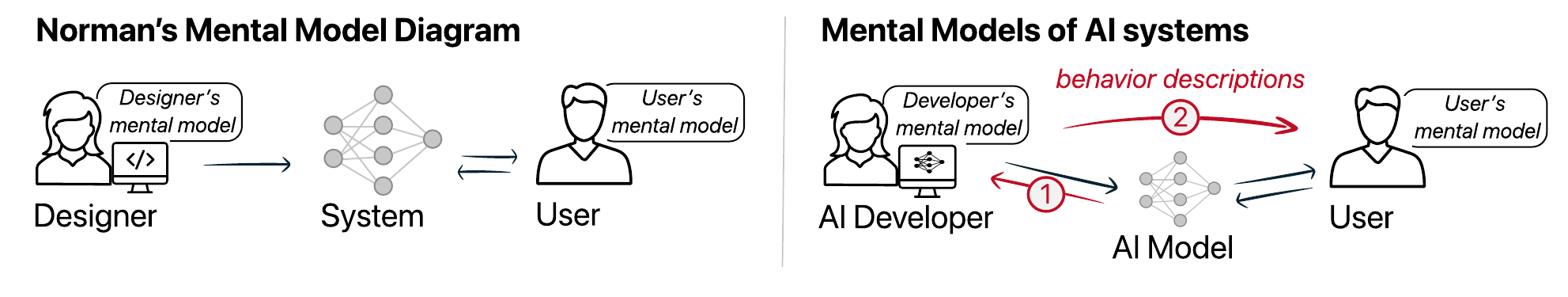}
    \caption{Don Norman's mental model framework \cite{Norman1987} describes how designers use their mental models to implement systems.
    End-users then interact with the systems and develop their own mental models of how they believe the systems work.
    While this process is similar for AI models, a key difference is that an AI is not a direct representation of a developer's intent, but a stochastic model learned from data.
    This means that \textbf{(1)} AI developers \textit{themselves} have to make sense of what an AI system has learned through testing and iteration.
    Subsequently, they can encode these insights as \textbf{(2)} \textit{behavior descriptions}, details of how an AI performs on subgroups of instances, that can be shown to end-users to improve human-AI collaboration.
    }
    \label{fig:model}
\end{figure*}


Human-AI collaboration is finding real-world use in tasks ranging from diagnosing prostate cancer \cite{Cai2019} to screening calls to child welfare hotlines \cite{de-arteaga_case_2020, kawakami_improving_2022}. 
To effectively work with AI aids, people need to know when to either accept or override an AI's output.
People decide when to rely on an AI by using their \textit{mental models} \cite{Kulesza2012, Bansal2019}, or internal representations, of how the AI tends to behave: when it is most accurate, when it is most likely to fail, etc. 
A detailed and accurate mental model allows a person to effectively complement an AI system by appropriately relying \cite{Lee2004} on its output, while an overly simple or wrong mental model can lead to blind spots and systematic failures~\cite{Bansal2019, bucinca_trust_2021}.
At worst, people can perform worse than they would have unassisted, such as clinicians who made more errors than average when shown incorrect AI predictions \cite{jacobs_how_2021, Bussone2015}.

Mental models are an inherently incomplete representation of any system, but numerous factors make it especially challenging to develop adequate mental models of AI systems.
First, modern AI systems are often black-box models for which humans cannot see how or why the model made a prediction \cite{Rudin2019}.
Second, black-box models are also often stochastic and can provide different outputs for slightly different inputs without human-understandable reasons \cite{Akhtar2018}.
Lastly, people often expect AI models to behave as humans do, which often does not match their actual behavior \cite{Luger2016} and makes people unaware of how an AI may fail \cite{Kang2015, zhang_ideal_2021}. 
These factors make it challenging for people to develop appropriate mental models of AI systems and effectively rely on them to improve their decision making.


To help people collaborate more effectively with AI systems, we propose directly showing end-users insights of AI behavior (\cref{fig:model}).
We term these insights \textbf{behavior descriptions}, details of an AI's performance (metrics, common patterns, potential failures, etc.) on subgroups of instances (subsets of a dataset defined by different features, e.g. ``images with low exposure'').
Behavior descriptions can take many forms, but should help end-users \textit{appropriately rely} \cite{Lee2004} on an AI, using its output when it is most likely correct and overriding it when it is most likely incorrect.
The goal of behavior descriptions is similar to that of explainable AI (xAI), but differs in what type of information is provided to end-users. 
Explainable AI attempts to describe \textit{why} an AI system produced a certain output, while behavior descriptions describe \textit{what} patterns of output a model tends to produce.
Thus, xAI and behavior descriptions can be used together to support effective human-AI collaboration.


We hypothesize that behavior descriptions will help people better detect systematic AI failures and improve AI-assisted decision making.
Additionally, we hypothesize that people will both trust an AI aid more and find it more helpful when shown behavior descriptions.
To test these hypotheses, we conducted human-subject experiments in three distinct domains: fake review detection, satellite image classification, and bird classification.
These three domains cover a range of data types, classification tasks, and human and AI accuracies to isolate the effect of behavior descriptions from domain-specific effects.

We found that behavior descriptions can significantly improve the overall accuracy of human-AI teams through two distinct mechanisms. 
First, for instances with a behavior description, users can directly correct AI failures.
Second, people tend to rely more on the AI system for instances without behavior descriptions when behavior descriptions are shown for other underperforming subgroups.
We additionally found that showing behavior descriptions had no significant impact on people's qualitative judgements, such as trust and helpfulness, of the AI.
Despite the potential benefits of behavior descriptions, their effects are not universal and depend both on how obvious AI failures are to a person and on a person's ability to correct the output once they know the AI is wrong.

In summary, this work introduces \textbf{behavior descriptions}, details shown to people working with AI systems of how an AI performs (metrics, common patterns, potential failures, etc.) for specific subgroups of instances.
We show how behavior descriptions can improve the performance of human-AI collaboration in three human-subject experiments.
These results indicate that knowing the mental models of end-users in human-AI collaboration is essential to understand how a human-AI team will perform and which interventions and decision aids will be most successful.




\section{Background and Related Work}
We review three main areas of work related to both creating and applying behavior descriptions.
First, we explore existing research on understanding and improving human-AI collaboration.
Second, we discuss methods for improving end-user AI understanding using explainable and interpretable AI.
Lastly, we describe tools and visualizations for analyzing AI behavior that can be used to create behavior descriptions.

\subsection{Human Factors and AI Aids}

When a person interacts with an AI system, they develop a \textit{mental model} of how it behaves - when it performs well, when it fails, or when it has quirky results \cite{Kulesza2012}.
Mental models let people effectively work with AIs by helping them decide whether to rely on, modify, or override an AI's output.
Therefore, it is important for people to have adequate mental models of AI systems so they can appropriately rely on an AI and override it when it is likely to fail \cite{Lee2004}.
There are various methods for encouraging \textit{appropriate reliance} of AI systems, often called \textit{trust calibration} \cite{Tomsett2020} techniques.

Studies have explored what factors influence people's mental models of AI systems.
\citet{Kulesza2013} found that people with more complete mental models were able to collaborate more effectively with a recommendation system.
\citet{Bansal2019} focused on attributes of AI systems and found that systems with \textit{parsimonious} (simple), and \textit{non-stochastic} (predictable) error boundaries were the easiest for humans to work with.
Other factors such as stated and perceived accuracy \cite{kocielnik_will_2019}, confidence values \cite{zhang_effect_2020}, and model personas \cite{Khadpe2020} can also influence people's mental models and their reliance on AI.
Behavior descriptions can complement these existing findings on human-AI collaboration, further helping people better collaborate with AI aids.

Recent methods have explored improving people's mental models, including tutorials explaining a task \cite{Lai2020}, tuning a model to better match human expectations \cite{martinez_personalization_2020}, or adaptive trust calibration \cite{Okamura2020}.
Some methods for improving mental models, such as adaptive trust calibration, use model details such as calibrated confidence scores that can be used in conjunction with behavior descriptions.
The technique most similar to our work is \citet{mozannar_teaching_2021}'s exemplar-based teaching of model behavior.
Their method learns nearest neighborhood regions around model failures to help validate people's mental models.
Although similar, behavior descriptions are \textit{semantic}, high-level insights of model behavior discovered and validated by developers.
Example-based training can fill gaps that were not previously identified by behavior descriptions.


Like model tutorials and trust calibration, behavior descriptions aim to improve human-AI collaboration by updating people's mental models of AI behavior.
Behavior descriptions provide additional information to end-users about \textit{when} to rely on an AI, further enriching their mental models and complementing these existing techniques.

\subsection{Explainable and Interpretable AI}
AI systems can be designed and explained to help people understand model outputs.
For example, interpretable model architectures allow users to inspect how a model makes predictions, while explanations can surface factors that affect model behavior.

One approach to improving end-user model understanding is to design inherently interpretable, or ``glass-box'', models.
Glass-box models can be easier to debug and trace, improving people's ability to understand AI predictions \cite{Rudin2019}.
Despite their benefits, glass-box models can be similarly difficult to collaborate with, for example, a study by \citet{poursabzi-sangdeh_manipulating_2021} found that more ``clear'' models can actually lead to information overload and hinder people's ability to detect AI errors.
Additionally, glass-box models may not work for more complex domains or data types such as images and videos.

Instead, another approach is model-agnostic, post-hoc explanations of model outputs.
LIME \cite{Ribeiro2016} is one such method, which fits linear models to a neighborhood of instances to highlight which features most influenced a model's prediction.
Instead of using all input features to explain a prediction, follow-up techniques such as Anchors \cite{Ribeiro2018} and MUSE \cite{lakkaraju_faithful_2019} learn sets of rules on features that are sufficient to explain predictions for a subset of instances.
Behavior descriptions also show details about a model for subsets of instances, but instead of approximating \textit{why} a model produces certain outputs, focuses on describing \textit{what} the pattern of outputs is.

Despite the insights post-hoc model explanations can provide end-users, they have been found to have a small or even detrimental effect on the performance of AI-assisted decision making \cite{Lai2019, chu_are_2020}. 
Explanations have been shown to potentially increase people's reliance on an AI, whether it is right or wrong, leading to worse team performance than a human or AI alone \cite{Bansal2020, Kaur2020, zhang_effect_2020, yang_how_2020}. 
By telling users how a model performs, behavior descriptions specifically target people's mental models and inform users when to rely on or override an AI's prediction.


Behavior descriptions can be combined with existing techniques to improve human-AI collaboration.
For example, behavior descriptions can still be used with interpretable models or shown in conjunction with black-box explanations.


\subsection{Behavioral Analysis of AI Systems}\label{sec:analysis}

There are numerous tools and techniques that help developers discover and fix model failures, especially those with real-world effects such as safety concerns and biases.
Behavioral analysis enables developers to go beyond aggregate metrics, such as accuracy, to discover specific and important behaviors \cite{Rahwan2019, cabrera_what_2022}.

Interactive tools, visualizations, and algorithmic techniques have shown promise in helping developers discover AI errors.
The tools most relevant to this work are focused on exploring model performance on a subset of data.
These include systems such as FairVis \cite{Cabrera2019}, a visual analytics system for discovering intersectional biases, AnchorViz \cite{chen_anchorviz_2018}, a visualization for semantically exploring datasets, and Slice Finder \cite{chung_slice_2019}, a method for automatically surfacing subsets of data with large model loss.
There are also tools tailored to specific domains, for example, Errudite \cite{Wu2019}, a visual analytics system for discovering and testing behaviors in NLP models.
More recently, crowd-based systems such as Pandora \cite{Nushi2018}, Deblinder \cite{Cabrera2021Deblinder}, and P-BTM \cite{Liu2020} have shown how crowdsourcing can augment behavioral analysis tools by surfacing previously unknown model failures.
Lastly, there are algorithmic testing methods for discovering AI behaviors.
A common technique is \textit{metamorphic testing} \cite{he_structure-invariant_2020}, which verifies the relationship between changes to an input and expected changes in an output.
Checklist \cite{Ribeiro2020} is one such testing tool for NLP models that creates sets of modified instances to test if language models understand basic logic and grammar. 

These model debugging techniques can be used to create behavior descriptions, statements of model performance on subgroups of data.
Additionally, analyses can be rerun on updated models to update behavior descriptions and directly inform end-users how a model has changed \cite{bansal_updates_2019}.

\section{Behavior Descriptions}

We define \textbf{behavior descriptions} as details of how an AI performs (metrics, common patterns, potential failures, etc.) for a subgroup of instances. 
Behavior descriptions should be semantically meaningful and human-understandable but can vary significantly between datasets and tasks.
For domains like binary classification, e.g. spam detection, they can be simple statements of accuracy like \textit{our system incorrectly flags marketing emails as spam 53\% of the time.}
In more complex domains like image captioning, they can describe specific behaviors like \textit{our system often describes mountain climbing as skiing} or \textit{our system can produce run-on sentences.}
The goal of behavior descriptions is to help end-users develop better mental models of an AI and thus \textit{appropriately rely} on the model --- using the AI output when it is more accurate and overriding the AI when it is more likely to fail.


In addition to what behavior descriptions contain, \textit{how} they are presented to end-users can vary and impact their effectiveness.
For example, \textit{when} behavior descriptions are shown can vary, from only showing them for extreme edge cases to every instance.
Similarly, behavior descriptions may only be shown during training or for the first few uses of an AI aid.
Due to the broad diversity of potential behavior descriptions, we do not attempt to enumerate all possible forms, and focus instead on testing some core assumptions of their efficacy.

\subsection{Principles for Effective Behavior Descriptions} \label{sec:props}

To design the behavior descriptions used in this work, we drew from existing studies of human-AI collaboration.
From these studies, we derived three properties behavior descriptions should have to maximize their effectiveness.
These are not the only principles, but an initial set used to design the behavior descriptions used in this study.

The first property comes directly from the goal of behavior descriptions, to help users appropriately rely on AI output.
Therefore, behavior descriptions should provide information that helps end-users decide both \textit{whether} they should rely on an AI and, if the AI is wrong, \textit{how} they should override it.
We summarize this principle as \textbf{actionable} behavior descriptions that provide end users with concrete information they can act on when using an AI aid.

The second principle comes from studies of AI error boundaries and explainable AI.
\citet{Bansal2019}'s study of people's mental models of AI systems found that models with errors that were \textit{parsimonious} (e.g. simple to define) led to more effective mental models.
Separately, \citet{poursabzi-sangdeh_manipulating_2021} found that showing end-users more details about an AI reduced their ability to detect AI errors due to information overload.
Thus, behavior descriptions should aim to be \textbf{simple}, short, and easy to interpret and remember.

The third and final principle comes from findings on alert fatigue and cognitive load.
When alerts or messages are shown continuously, people can suffer from \textit{alert fatigue} and begin to ignore the messages \cite{cash_alert_2009}.
Additionally, showing people more information increases their cognitive load, which can lead to decreased learning and performance \cite{van_gog_effects_2011, sweller_element_2010}.
To avoid these pitfalls, behavior descriptions should be limited and focus on the subgroups of instances with the highest impact.
Thus, the third principle is to aim for \textbf{significant} behavior descriptions, either common behaviors or those with the most serious consequences.

In summary, the three design principles that we followed to create the behavior descriptions in this work are the following:

\begin{enumerate}
    \item \textbf{Actionable}, suggesting both whether and how a person should override an AI output.
    \item \textbf{Simple}, aiming to be as parsimonious and easy to remember as possible.
    \item \textbf{Significant}, limited to behaviors that are common and/or have serious consequences.
\end{enumerate}




\subsection{Why Not Just Fix AI Failures?}

A key question surrounding the utility of behavior descriptions is why developers would not directly fix the systematic model failures or behaviors they discover. 
Modern AI systems are often stochastic, black-box models like neural networks that cannot be deterministically ``fixed'' or ``updated''.
ML practitioners interviewed by \citet{holstein_improving_2019} would often try to fix one problem that would cause the model to start failing in other unrelated ways.
In another empirical study, \citet{Hopkins2021} found that practitioners would avoid or limit model updates due to concerns about breaking their models and introducing more failures. 
Challenges to fixing known model failures are also present in research.
This is most apparent with natural language processing models, which are approaching human aptitude in many domains such as question answering and programming \cite{Brown2020}.
Despite the growing capability of NLP models, many state-of-the-art systems still encode serious biases and harms \cite{Caliskan2017} and fail basic logical tests \cite{Ribeiro2020} that developers have been aware of for years, but have not been able to fix.

Fixing model failures can also require significant amounts of new training data, which tends to be expensive and time consuming.
Additionally, the specific instances needed to improve certain model failures can be difficult to get, such as globally diverse images or accents \cite{Kuznetsova2020}.
Behavior descriptions can be an important intermediate solution for model issues; end-users can effectively work with an imperfect AI that developers are working to improve and fix.

Lastly, behavior descriptions can be helpful when models are updated.
Model updates can be incompatible with end-users' existing mental models and violate their expectations of how an AI behaves, leading to new failures \cite{bansal_updates_2019}.
Updated behavior descriptions can be deployed with a new model to directly update end-user's mental models and avoid decreased performance.

\section{Experimental Design}

To directly test how behavior descriptions impact human-AI collaboration, we conducted a set of human-subject experiments across three different tasks.
The tasks range across dimensions such as human accuracy, AI accuracy, and data type to reduce the chance that our results are confounded by domain-specific differences.

\subsection{Experimental Setup}

To test the effect of behavior descriptions, we conducted experiments across three different classification tasks.
All three tasks shared the same core setup and only varied in the type of classification task (binary or multiclass) and what participants were asked to label.
For each task, we tested three between-subjects conditions to isolate the effect of behavior descriptions:

\begin{itemize}
    \item \textbf{No AI:} Participants were asked to classify instances without any assistance.
    \item \textbf{AI:} Participants were asked to classify instances with the help of an AI they were told was 90\% accurate.
    \item \textbf{AI + Behavior Descriptions (BD):} Participants were asked to classify instances with the help of an AI they were told was 90\% accurate.
    Additionally, for instances that were part of a behavior description group (10/30 instances), participants were shown a behavior description stating the AI accuracy for that type of instance (see \cref{sec:groups} for details about behavior description groups).
\end{itemize}

Participants in every condition and task were first shown a consent form, introduced to the task, and shown example instances and labels. 
Those in the condition with the AI were also shown a screen before labeling that introduced the AI and stated its overall accuracy of 90\%. 
The participants were then shown and asked to label 30 instances (see \cref{fig:ui} for an example UI), 20 instances from the overall dataset, and 5 instances each from two subsets of the data, \textit{behavior description groups}, where the AI performance was significantly worse than for the overall model (see \cref{sec:groups} for details).
After labeling the 30 instances, participants completed a short questionnaire with Likert scale questions, open-ended text responses, and an attention check question. 


The experiments were conducted on Amazon Mechanical Turk (AMT) with participants from the United States.
We selected participants who had completed more than 1,000 tasks and had an approval rating of more than 98\% to ensure high-quality responses. 
Participants that failed the attention check, a question asking which step they were currently on, were still paid, but were excluded from the results and analysis. 
Although we considered providing bonuses as an incentive for accurate responses, we found that the incentive to have the task approved was sufficient to get good results without a bonus.
We confirmed this by finding similar average accuracy on the control condition for the reviews task, 56\%, to that reported by \citet{Lai2019} on the same task using a bonus, 51\%.

\begin{figure*}
    \centering
    \includegraphics[width=\textwidth]{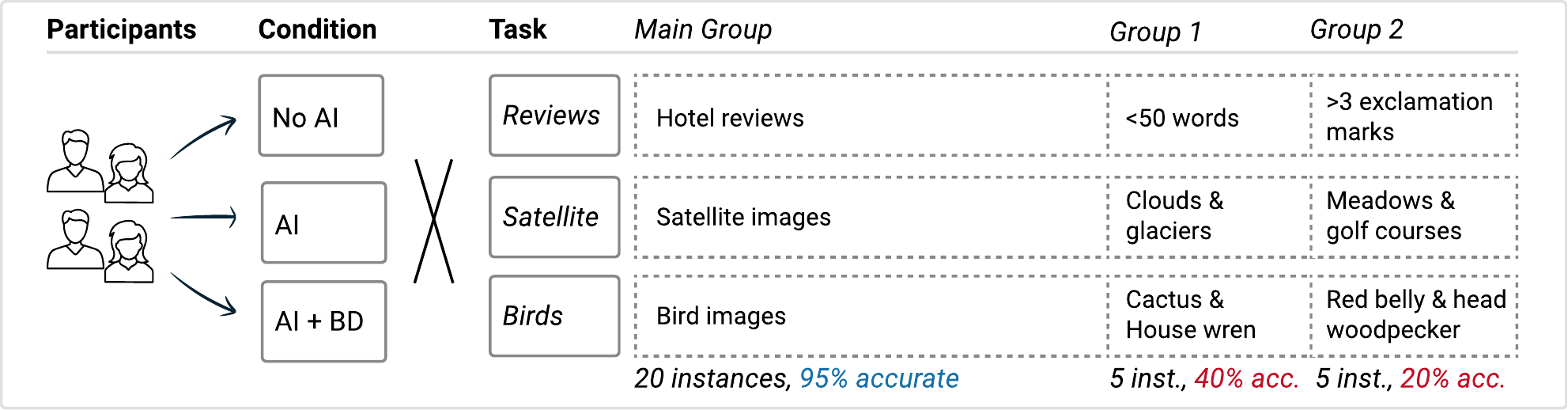}
    \caption{\textbf{Experimental setup}. 
    Each participant was randomly assigned to a condition and dataset (3x3 between-subjects study, 25 participants per condition, 225 participants total). 
    For each dataset, participants saw 30 instances, 20 instances from the whole dataset with an AI accuracy of 95\%, and 5 instances each from two subsets of the dataset with an AI accuracy of 40\% and 20\% respectively (simulating subgroups behavior descriptions would be useful for).
    In the \textit{AI + Behavior Description} condition, participants were shown behavior descriptions for instances in group 1 and 2.
    While the instances shown to participants were randomly chosen from a larger subset of data, each participant saw the same number of AI errors to ensure they observed the same AI accuracies.
    }
    \label{fig:conditions}
\end{figure*}

The study was approved by an Institutional Review Board (IRB) process.
We had a total of 225 participants, 25 per task/condition pair.
The number of participants per condition was chosen using a power analysis on the mean and standard error of the accuracy in the initial usability studies for the interface.
Of the 225 participants, we removed 13 for failing the attention check.
Additionally, we removed three other participants, across two conditions, who had an accuracy of less than 10\% (the next highest accuracy being 35\%), the same as guessing randomly.
We paid participants \$2 for the task, which lasted 15 minutes for an hourly compensation of \$8 an hour.

\subsubsection{Behavior descriptions and wizard-of-oz AI}\label{sec:groups}

For this work, we used behavior descriptions stating the model accuracy for subgroups of the data for which the model performs significantly worse than average \cite{Cabrera2019,Nushi2018,chung_slice_2019,Ribeiro2020,wu_local_2019}.
Depending on whether the task was binary or multiclass classification, the behavior descriptions resembled text sentences such as "the model is 20\% accurate for this type of instance" or "the model confuses these two classes 80\% of the time".
These types of behavior descriptions are straightforward to create, calculating accuracy on a subset of data, and are actionable for end-users, informing them of how likely it is that they need to override the AI.

In order to control the distribution of instances and AI accuracy each participant saw, we used a wizard-of-oz AI system, mock AI outputs, with a fixed observed accuracy.
Of the 30 instances each participant labeled, 20 were randomly chosen from the overall dataset, and the final 10 split into two groups of 5 instances randomly selected from two subsets of the dataset which we term \textit{behavior description groups}.
The AI accuracy for instances in the behavior descriptions groups was significantly lower than the overall model to simulate the types of subgroups behavior descriptions would be used for.
Lastly, while we fixed the distribution of correct/incorrect model outputs per behavior description group, the instances each participant saw were randomized, both sampled from a larger set of images and randomly ordered.

The accuracy breakdown for the subgroups was the following: 95\% accuracy (1/20 misclassified) for the \textit{main group} of 20 instances, 40\% accuracy (3/5 misclassified) for the first behavior description group, \textit{group 1} and 20\% accuracy (4/5 misclassified) for the second group, \textit{group 2}.
The behavior description groups for each task are detailed in \cref{fig:conditions} and \cref{sec:tasks}.
The purpose of the behavior description groups is to have two concrete subsets representing the type of instances for which a behavior description would be used - these are the instances for which we show behavior descriptions in the AI + BD condition.

\begin{figure*}
    \centering
    \includegraphics[width=\textwidth]{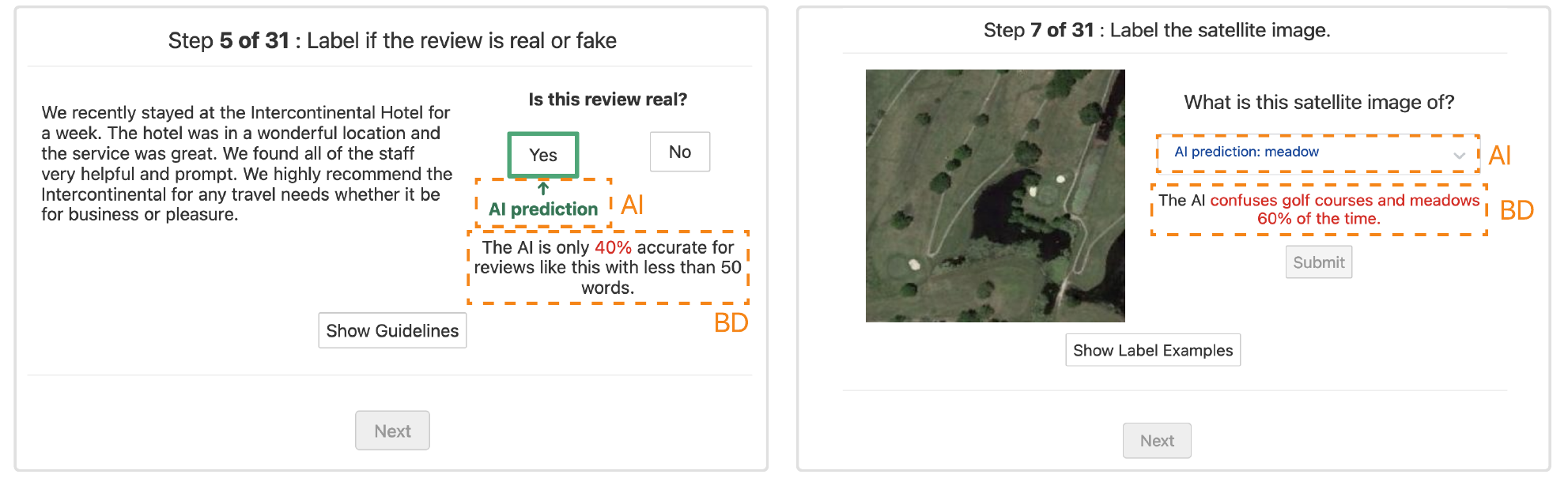}
    \caption{
    UI screenshots for the fake reviews (left) and satellite image classification (right) tasks. 
    Each participant labeled 30 instances, distributed according to the instance groups described in Figure \ref{fig:conditions} and Section \ref{sec:tasks}.
    Both screenshots are shown on a labeling step for the \textit{AI + Behavior Description (BD)} condition and on instances that are part of a behavior description group.
    In the \textit{AI} condition participants are not shown the additional text for instances in a BD group, and in the \textit{No AI} condition participants are not shown the AI output.
    The bird classification task used the same format as the satellite classification task shown.}
    \label{fig:ui}
\end{figure*}

The accuracy breakdown above gives an actual AI accuracy of 73.33\%, not the 90\% accuracy stated to the participants, since the task would have been too long to have both an actual 90\% overall accuracy and significantly low accuracies for the two behavior description groups.
We wanted to simulate a situation where an AI is reasonably accurate, e.g. > 90\%, so that a human would want to work with it.
Existing work has explored the impact of stated versus observed accuracy and found that there is a small decrease in agreement with the AI the lower the observed accuracy \cite{yin_understanding_2019}.
Since each condition had the same stated vs. observed accuracy discrepancy (90\% vs. 73.33\% respectively), it should not impact our relative findings on the efficacy of behavior descriptions.


\subsubsection{Classification tasks}\label{sec:tasks}

We chose three distinct tasks for the study to ensure that our findings are not tied to a specific dataset or task.
The three tasks vary by data type (text/image), task type (binary/multiclass classification), and human accuracy (human better/worse than AI).
For each task description below, we also detail what types of instances make up \textit{group 1} and \textit{group 2}, the subsets of instances (behavior description groups) for which the AI is less accurate and for which behavior descriptions are shown in the BD condition (see \cref{fig:conditions}).
The tasks are the following:

\begin{enumerate}[leftmargin=0cm,itemindent=.5cm,labelwidth=\itemindent,labelsep=0cm,align=left]
    \setlength\itemsep{0.5em}
    \item[] \noindent\textbf{Fake Review Detection.} The dataset of deceptive reviews from \citet{Ott2011, Ott2013} has 800 truthful reviews and 800 deceptive reviews for hotels in Chicago.
    The truthful reviews were collected from online sites such as TripAdvisor, while false reviews were collected from Amazon Mechanical Turk workers.
    The objective of the task is to determine whether the review is ``truthful'', written by someone who actually stayed at the hotel, or ``deceptive'', written by someone who has not. 
    We chose this task since it has been used in previous studies of human-AI collaboration to test the effect of explanations~\cite{Lai2019} and tutorials~\cite{Lai2020}.
    BD groups were chosen from research on common failures in language models:
    \\
    \textit{Group 1} - Reviews with less than 50 words \cite{Ribeiro2020}.
    \\
    \textit{Group 2} - Reviews with more than 3 exclamation marks \cite{Teh2015}.

    \item[] \noindent\textbf{Satellite Image Classification.} The satellite images come from the NWPU-RESISC45 dataset \cite{Cheng2017}, which has 31,500 satellite images across 45 classes.
    The task for the dataset is multiclass classification, labeling each square satellite image with a semantic class.
    We selected a subset of 10 classes for the task in order to show participants at least a couple instances per class.
    This task was inspired by real-world human-AI systems for labeling and segmenting satellite images \cite{logar_pulsesatellite_2020}.
    The BD groups were chosen from areas of high error in the confusion matrices from the original paper \cite{Cheng2017}:
    \\
    \textit{Group 1} - Cloud and glacier images
    \\
    \textit{Group 2} - Meadow and golf course images

    \item[] \noindent\textbf{Bird Classification.} The bird images came from the Caltech-UCSD Birds 200 dataset \cite{WelinderEtal2010}, made up of 6,033 images of 200 bird species.
    As in the satellite image task, we chose a subset of 10 classes from the dataset for multiclass classification.
    The task was inspired by numerous apps and products for classifying bird species \cite{van_horn_building_2015}.
    The BD groups were chosen from birds in the same family, classes that are the most similar and difficult to distinguish:
    \\
    \textit{Group 1} - Cactus Wrens and House Wrens
    \\
    \textit{Group 2} - Red Bellied Woodpeckers and Red Headed Woodpeckers

\end{enumerate}

\subsection{Hypotheses}

From the primary goal of behavior descriptions, helping end-users appropriately rely on an AI, we formulated the following hypotheses of how we expect behavior descriptions to affect human-AI collaboration.
The hypotheses focus both on quantitative measures of performance and qualitative opinions from participants.
Our first hypothesis is that behavior descriptions will improve the overall accuracy of human-AI teams.
We hypothesize that behavior descriptions will help end-users more appropriately rely on the AI \cite{Lee2004}, leading to improved performance.

\begin{itemize}
    \item[\textbf{H1.}] Showing participants behavior descriptions (BD) results in higher overall accuracy than just showing the AI prediction or no AI. 
\end{itemize}

\noindent We hypothesize that this improved performance will primarily come from participants overriding the systematic failures identified by behavior descriptions.
By providing actionable descriptions of when the model is most likely to be wrong, for instances in BD groups, we hypothesize that the participants will be able to better identify and correct errors when shown behavior descriptions. 

\begin{itemize}
    \item[\textbf{H2.}] The higher accuracy from showing BDs is due to a higher accuracy on instances that are part of BD groups.
\end{itemize}

\noindent Lastly, we have a set of hypotheses on how we expect people's perception of the AI to change when they are shown behavior descriptions.
We hypothesize that participants will find the AI to be \textit{more helpful}, will be \textit{more likely to want to use the AI}, and will \textit{trust the AI more} when they are shown behavior descriptions.
These hypotheses come from \citet{yin_understanding_2019}'s study on accuracy and trust in AI systems, which found that observed accuracy significantly affected trust and reliance on AI systems.

\begin{itemize}
    \item[\textbf{H3a.}] Participants shown BDs trust the AI more than when just shown the AI output.
    \item[\textbf{H3b.}] Participants shown BDs find the AI more helpful than when just shown the AI output.
    \item[\textbf{H3c.}] Participants shown BDs are more likely to want to use the AI in the future than when just shown the AI output.
\end{itemize}



\section{Results}

To assess the significance of different conditions on participant accuracy, we used ANOVA tests with Tukey post-hoc tests to correct for multiple comparisons.
For the Likert scale questions, we used Mann-Whitney U tests with Bonferroni corrections.
Lastly, we used linear models to test for learning effects, also using a Bonferroni correction for multiple comparisons.
We used a $p$ value of 0.05 as the cutoff for significance.

\subsection{Overall Accuracy}

\begin{figure*}
    \centering
    \includegraphics[width=\textwidth]{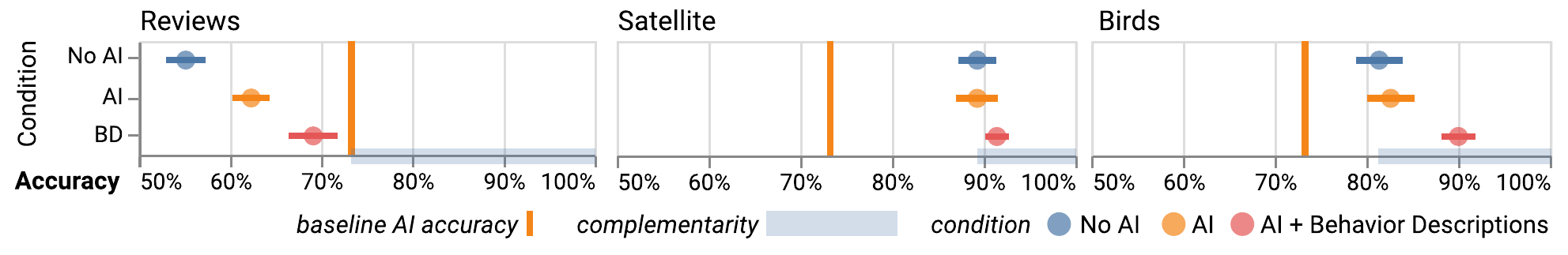}
    \caption{\textbf{Average participant accuracy by task and condition.} 
    The vertical orange bar indicates the AI accuracy, what would be the participant's accuracy if they picked the AI response every time. 
    The blue shaded area indicates \textit{complementarity}, the region where the human+AI accuracy is higher than either the human or AI alone. 
    We find that behavior descriptions led to higher accuracy in the reviews and birds tasks, with complementarity in the birds task (red point in rightmost chart).
    The error bars represent standard error.
    }
    \label{fig:acc-results}
\end{figure*}

To test \textbf{H1} we can compare the average human-AI team accuracy for each task across the three conditions.
Overall, we found that the \textit{AI} and \textit{AI + behavior descriptions (BD)} interventions have a different effect on team performance in each task (\cref{fig:acc-results}).

For the reviews task, there was a significant difference in accuracy across the three conditions $(F_{2, 64}=9.18, \ p<0.001)$.
The only significant pairwise difference was between the \textit{No AI} and \textit{AI + BD} conditions $(p < 0.001, \ 95\% \ C.I. = [0.06, 0.22]$).
While the AI by itself did not significantly improve the accuracy of the participants, AI supplemented with behavior descriptions led to significantly higher team accuracy, supporting \textbf{H1}.
Despite the increased performance, there was no human-AI complementarity -- higher accuracy than either the human or AI alone -- as participant accuracy at every condition was lower than the baseline AI accuracy.

In the satellite classification task there was no significant difference between conditions $(F_{2, 67}=0.63, \ p=0.534)$.
The baseline human accuracy without AI support was the highest across tasks, around 90\%, so there was a smaller margin to improve the accuracy of the participants using an AI with a significantly worse accuracy.

Lastly, in the birds classification task, there was a significant difference in participant accuracy $(F_{2, 65}=3.98, \ p=0.023)$.
As in the reviews task, the only pairwise difference was between the \textit{No AI} and \textit{AI + BD} conditions  $(p = 0.048, \ 95\% \ C.I. = [0.00, 0.16]$), showing how behavior descriptions can lead to significant increases in participant accuracy using an AI and supporting \textbf{H1}.
This increased performance also led to complementary human-AI accuracy, higher than that of both the AI or human alone.

In sum, these results \textbf{partially support H1}, with behavior descriptions leading to significantly higher accuracy in two of the three tasks.
This suggests that while behavior descriptions will not universally improve the accuracy of human-AI teams, they can lead to significant improvements in certain tasks and domains.

\begin{figure*}
    \centering
    \includegraphics[width=\textwidth]{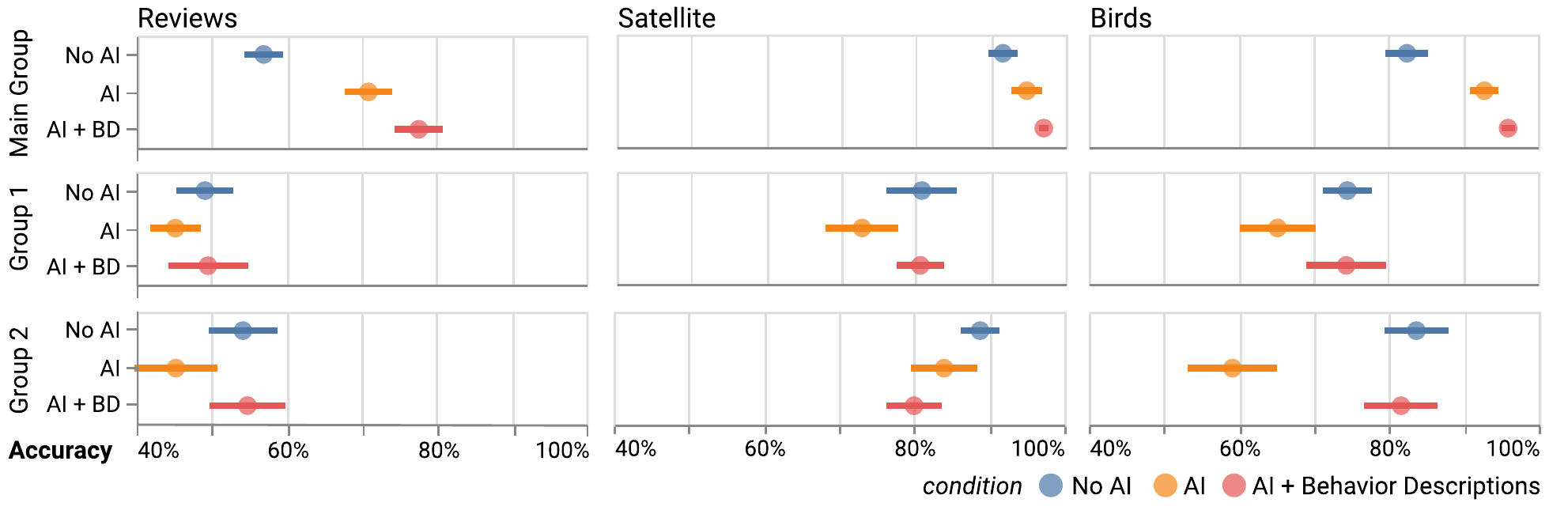}
    \caption{\textbf{Average team accuracy by task, condition and instance group.} 
    We further break down accuracy by instance type (see \cref{sec:groups}): the main group (20 instances), and two behavior description groups (5 instances each).
    The average human-AI accuracy across the three groups of instances gives us an idea of \textit{how} behavior descriptions improve the performance of human-AI teams.
    We find that participants relied more on the AI when shown BDs in every task.
    Participant performance on the different behavior description groups was mixed, from no effect to significant improvement in group 2 birds (bottom right).
    These results highlight the two effects of behavior descriptions, increasing human reliance on a more accurate AI and overriding systematic AI errors.
    The error bars represent standard error.
    }
    \label{fig:groups}
\end{figure*}

\subsection{Accuracy by Behavior Description Group}

To better understand the ways in which behavior descriptions impact performance, we can look at participant accuracy across both conditions and instances in the three different behavior description groups described in \cref{sec:groups} (\cref{fig:groups}).
This allows us to directly test \textbf{H2} by seeing if participants in the \textit{AI + BD} condition perform significantly better on the two subsets of instances that have behavior descriptions.
The results can be seen in \cref{fig:groups}.
We found that this is partially true, as the increased accuracy of the \textit{AI + BD} condition was due both to higher accuracy on the behavior description groups \textit{and} instances in the main group.

In the reviews task, there was only a significant difference between conditions for instances in the normal group $(F_{2, 64}=11.82, \ p<0.001)$, with no differences between conditions for either of the behavior description groups.
The \textit{AI} $(p = 0.007, \ 95\% \ C.I. = [0.03, 0.24]$) and \textit{AI + BD} $(p < 0.001, \ 95\% \ C.I. = [0.10, 0.31] $) conditions were significantly more accurate than the \textit{No AI} condition for instances in the main group.
These results do not support \textbf{H2}, as the higher overall accuracy of the \textit{AI + BD} condition was primarily due to greater reliance on the AI for instances in the main group, not higher accuracy on instances with behavior descriptions.

Despite there being no difference in overall accuracy between conditions for the satellite task, there were differences when looking at the behavior description groups.
As with the reviews task, there was a significant difference between conditions for instances in the main group $(F_{2, 67}=5.34, \ p=0.007)$.
Participants in both the \textit{AI} $(p = 0.026, \ 95\% \ C.I. = [0.00, 0.09]$ and \textit{AI + BD} $(p = 0.007, \ 95\% \ C.I. = [0.01, 0.10]$ conditions had a higher accuracy than participants in the \textit{No AI} condition for instances in the main group.
This is the same result we found in the reviews task, where there were significant accuracy differences for instances in the main group.
Despite this difference, the increased accuracy did not translate to a higher overall accuracy for the \textit{AI} and \textit{AI + BD} conditions.
Since we did not find any difference between conditions for the two BD groups, these findings do not support \textbf{H2}.

Lastly, we found significant differences between conditions for multiple behavior description groups in the bird classification task.
As with the other two tasks, there is a significant difference in accuracy between conditions for instances in the main group $(F_{2, 65}=12.22, \ p<0.001)$.
Both \textit{AI} and \textit{AI + BD} have a significantly higher accuracy than \textit{No AI}, but there is no significance between \textit{AI} and \textit{AI + BD}.
Although there is no significance for instances in group 1, we do see a difference between conditions for instances in group 2 $(F_{2, 65}=6.81, \ p=0.002)$. 
Both the \textit{No AI} and \textit{AI + BD} conditions both have a higher accuracy than just \textit{AI}. 
This shows that while participants were able to distinguish between the two types of woodpeckers in group 2 without the AI or when informed about the AI failures, participants trusted the AI and performed significantly worse when just shown the AI prediction.
These results support \textbf{H2}, as the increased accuracy on group 2 led to higher accuracy in the AI + BD condition.

In summary, these results \textbf{partially support H2}.
Depending on the task, the higher accuracy in the \textit{AI + BD} condition was due to a higher accuracy on instances that were part of BD groups \textit{and} and a greater reliance on the AI for instances in the main group.

\subsection{Qualitative Results}

\begin{figure*}
    \centering
    \includegraphics[width=\textwidth]{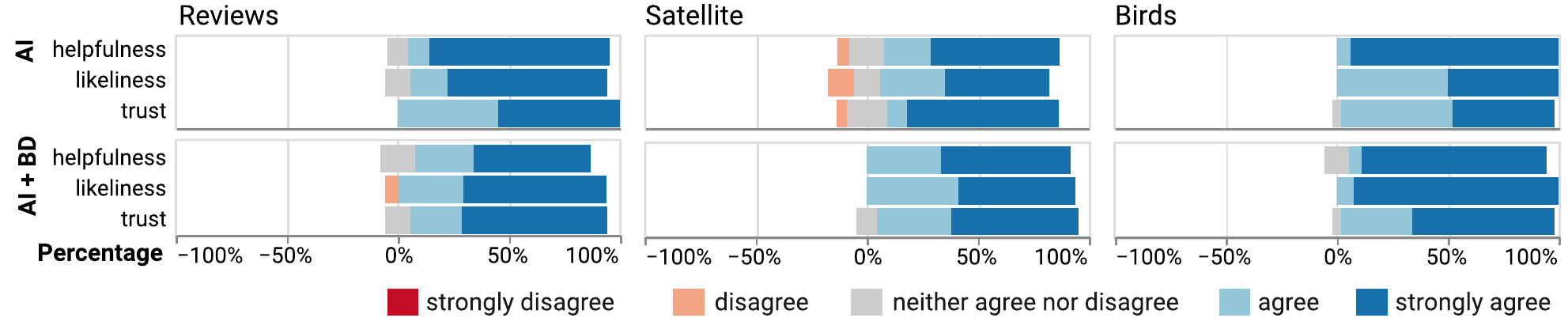}
    \caption{\textbf{Likert-scale responses on perception of AI.} 
    The diverging stacked bar chart centered around the neutral response shows that participants across all conditions and subjective measures overwhelmingly viewed the AI favorably.
    There were no significant differences in user's perception of the AI when they were give behavior descriptions.
    }
    \label{fig:likert}
\end{figure*}

After labeling the 30 instances, participants were shown a questionnaire page asking about their opinions and feelings of the AI aid using Likert scale questions (\cref{fig:likert}).
Since these questions were directly related to the AI output, they were only shown to participants in the \textit{AI} and \textit{AI + BD} conditions.
The questions were the following: (1) How \textbf{helpful} was the AI for this task? (2) How \textbf{likely} are you to use this AI again in a future task? (3) How much do you \textbf{trust} the AI?
We did not find significant differences between the \textit{AI} and \textit{AI + BD} conditions for any of the Likert scale questions across tasks.
These results reject \textbf{H3a, H3b, H3c}, and indicate that behavior descriptions do not significantly impact participant perceptions of AI despite differences in how participants use AI predictions.

\subsection{Additional Findings}

In addition to the accuracy metrics and Likert scale responses, we collected and analyzed additional measurements and qualitative responses to further unpack the dynamics of behavior descriptions.
Although these are post hoc, exploratory results for which we did not have hypotheses, they can serve as inspiration for further, more formal studies.

One such factor was the time it took participants to label each instance, which can potentially surface interesting insights when compared between conditions and behavior description groups.
Unfortunately, the time per instance had high variance and was inconsistent, with numerous outliers.
This is likely due to the way AMT workers complete tasks, as they often take breaks or search for new HITs while working on a task \cite{lascau_monotasking_2019}.
Future studies could incentivize quick responses to gather more accurate time data and measure the speed of participant responses across conditions.

\begin{figure*}
    \centering
    \includegraphics[width=\textwidth]{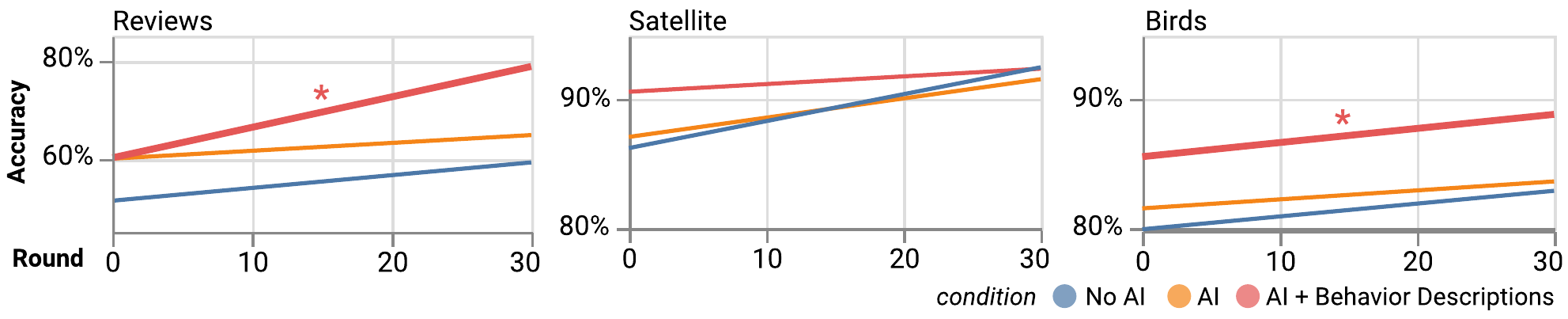}
    \caption{\textbf{Learning curves by task and condition.} 
    We fit linear models of accuracy on round number to measure learning effects.
    We found that the two conditions in which BDs significantly improved performance also had significant learning effects, the \textit{AI + BD} conditions in the reviews and birds tasks (denoted by *).
    }
    \label{fig:learning}
\end{figure*}

We also tracked in which round each instance was labeled, $(n/30)$, to detect any learning effects (\cref{fig:learning}).
To measure learning effects, we fit a linear model of average participant accuracy for each condition and each step.
We found that for the two domains in which behavior descriptions were effective, reviews and birds, there was also a significant learning effect for the \textit{AI + BD} condition (reviews/AI + BD: $\beta = 0.0062, p = 0.020$; birds/AI + BD: $\beta = 0.0031, p = 0.035$).

The endpoints of the learning curves also show some interesting patterns.
For the reviews task, participants in the \textit{AI} and \textit{AI + BD} conditions started at similar accuracies and only over time did participants in the \textit{AI + BD} condition learn to effectively work with the AI and make use of the behavior descriptions.
In contrast, in the birds classification task, participants in the \textit{AI + BD} condition were consistently better than the \textit{AI} condition and improved at a similar rate over time.
Interestingly, for the satellite domain, the \textit{AI + BD} condition learning curve ends at a point similar to the \textit{No AI} condition but starts significantly higher.
This suggests that, while over time participants learned the correct satellite labels or when the AI tended to fail, the behavior descriptions helped bootstrap the learning process.

Lastly, we collected qualitative free response answers from participants about patterns of AI behavior they noticed and general comments they had about the task.
As expected, participants in both the \textit{AI} and \textit{AI + BD} noticed that the AI failed in the behavior description groups.
While more participants described failures in the \textit{AI + BD} condition than the \textit{AI} condition, the comments were inconsistent and did not show many significant differences between conditions.
We found an interesting pattern from comments in the birds task, where participants in the \textit{AI} condition described more general but correct patterns of AI failure.
Specifically, a participant found that \textit{``the AI is good at predicting the main class of birds, but might get the sub class incorrect,''} which was the common failure reason between the two BD groups.

\section{Discussion}

These results show that directly informing people about AI behaviors can significantly improve human-AI collaboration.
We hope these initial insights spark future work on understanding people's mental models and developing new types of behavior-based decision aids.

\subsection{Effectiveness of Behavior Descriptions}

Overall, the results generally supported our primary hypothesis that behavior descriptions can significantly improve the accuracy of participants working with AI aids.
Surprisingly, however, we found that the improved accuracy came from two complementary effects: people overriding systematic AI failures \textit{and} relying more often on the AI for instances without behavior descriptions.

The first effect, helping participants fix systematic AI failures, was the initial goal of behavior descriptions, but we found that it was inconsistent and varied significantly between conditions and behavior description groups.
For example, in the reviews task, there was no significant accuracy difference for instances in either behavior description group.
This is likely due to the behavior descriptions in the reviews domain not being sufficiently \textit{actionable} and the participants not knowing whether or not to override the AI when they knew it was likely wrong.
On the contrary, there \textit{was} a significant difference in accuracy for instances in group 2 of the bird classification task, where participants in the BD condition were able to fix AI failures that most participants with just the AI did not notice.
Thus, while we did find some support for this effect, participants fixing AI failures in BD groups was often not the main driver of increased overall accuracy.

In contrast, the more consistent effect was participants in the \textit{AI + BD} condition relying on the AI more often for instances not in the BD groups (main group) compared to the \textit{AI} condition.
This increased reliance on the AI when it was more accurate, on instances in the main group, contributed to the higher overall accuracy in the \textit{AI + BD} conditions for the reviews and birds tasks.
This was an unexpected effect which was a factor in the higher overall accuracy for participants in the \textit{AI + BD} condition.
Although unexpected, the effect is corroborated by studies on trust in AI systems that found that low reliability, AI output that violates people's expectations over time, decreases trust and reliance in an AI \cite{Glikson2020}.
By isolating common AI errors into well-defined, predictable subgroups, the AI appears more reliable to people and can increase their trust and reliance.

Although two of the three domains showed an overall increase in accuracy with behavior descriptions, there was no significant increase in the satellite classification task.
We believe that this is due to the high baseline accuracy of humans performing the task, with approximately 90\% accuracy, which left little room for improvement with a significantly less accurate AI.
The high accuracy of the participants also allowed them to detect and fix the AI errors in the BD groups without any prompting in the \textit{AI} condition.
In summary, while behavior descriptions can improve performance, they are not the panacea to human-AI collaboration --- AI aids must still provide value and complementarity to human decision makers.

\subsection{Learning and Behavior Descriptions}

We found that participants improved more quickly when using behavior descriptions, learning to effectively complement the AI.
The primary pattern we noticed was the significant learning rate in the \textit{AI + BD} condition for the reviews and birds task, as participants quickly learned when they should override the AI.
This could be an important property for AI systems that are updated often, as new behavior descriptions could be used to directly update end-users mental models and avoid failures from incompatible updates \cite{bansal_updates_2019}.

A secondary effect we found was a higher initial accuracy in the \textit{AI + BD} condition for the satellite and birds task.
While it took time for participants in the \textit{AI} condition to notice how a model tended to fail, the participants with behavior descriptions were aware of the failures right from the start.
Even if people's accuracy converges over time, as in the reviews task, behavior descriptions can speed up end-user onboarding and improve early-stage performance.




\subsection{Authoring Behavior Descriptions}

The AI debugging techniques described in Section \ref{sec:analysis} can be used to create behavior descriptions, but tools designed specifically for creating BDs could provide important benefits.
For example, behavior descriptions do not \textit{have} to be AI failures, but could highlight consistent output patterns or areas where the model performs much better than humans.
Bespoke tools could also optimize for creating behavior descriptions with effective properties such as those described in Section \ref{sec:props}.
Tools customized to create behavior descriptions could optimize for these different properties.

The types of people who create behavior descriptions can also be much broader than AI/ML developers.
With the right tools, stakeholders ranging from quality assurance engineers to domain-specific teams could discover and deploy their own behavior descriptions.
In the future, techniques from crowdsourcing could be used to harness end-user's own insights for generating behavior descriptions.
This could be, for example, prompting end-users to report consistent failures and patterns that could then be aggregated and voted on \cite{Cabrera2021Deblinder}.
The insights could be processed to automatically generate up-to-date behavior descriptions.


\subsection{Understanding and Improving Mental Models of AI}

These experiments also implicitly tested the more general effect of human mental models on how they collaborate with AIs. 
We found that when humans have more detailed mental models of how an AI performs, they are more likely to rely on the AI in general.
This result is interesting regardless of the use of behavior descriptions, as more experienced end-users will likely develop better mental models of AI aids and gradually change how they rely on the AI.

Developing better techniques for quantifying end-user's mental models of AI systems can help researchers design effective decision aids such as behavior descriptions.
This work could take inspiration from studies in HCI and psychology on \textit{cognitive modeling}, mathematical models of human cognition \cite{anderson_act-r_1997}. 
Cognitive models have been used in education by simulating how people learn math to dynamically teach students \cite{ritter_cognitive_2007}.
For human-AI collaboration, cognitive models of how people learn AI behavior could inform the design of decision aids such as behavior descriptions.

Our results can also guide the design of machine learning models that can more directly provide behavior descriptions. 
For example, sparse decision trees could be used to directly generate behavior descriptions to show to end-users. 
New algorithmic methods that are more amenable to finding clusters of high error could lead to better human-AI collaborative systems.


\section{Limitations and Future Work}

The participants in our study were Amazon Mechanical Turk workers with limited domain expertise in the tasks they completed.
Domain experts and professionals, e.g., doctors or lawyers, have deeper expertise in their fields and may develop different mental models of AI systems they work with.
For example, they may notice AI failures more often or be less influenced by the information provided by behavior descriptions. 
Future studies can explore the dynamics of using BDs with domain experts.

Although we selected domains that reflect potential real-world examples of human-AI collaboration, our experiment was a controlled study in a simulated setting.
The impact of behavior descriptions may vary when applied to real-world situations with consequential outcomes \cite{zhang_effect_2020}, such as a radiologist classifying tumors.
When classification errors have a much higher cost, people may update their mental models differently or act more conservatively when shown BDs.

Our study used one specific type of behavior description, subgroup accuracy.
There are countless variations of BDs that can be explored further, such as highlighting subgroups with high accuracy, describing what features of an instance are correlated with failure, or suggesting alternative labels.
Future experiments could test these variations to disentangle which features of behavior descriptions are the most effective in improving people's performance.

Lastly, our study focused on the relatively simple domain of classification.
Modern AI systems are used in much more complex tasks such as image captioning, human pose estimation, and even image generation.
The behavior descriptions for these domains will likely look significantly different from those tested in this work.
For example, BDs for a captioning model might focus on grammatical issues and object references rather than statistical metrics.
The impact of behavior descriptions will likely vary significantly in these domains, and specific studies could explore both their effect and optimal designs.

\section{Conclusion}
We introduce \textbf{behavior descriptions}, details shown to people in human-AI teams of how a model performs for subgroups of instances.
In a series of user studies with 225 participants in 3 distinct domains, we find that behavior descriptions can significantly improve human-AI team performance by helping people both correct AI failures and rely on the AI when it is more accurate.
These results highlight the importance of people's mental models of AI systems and show that methods directly improving mental models can improve people's performance when using AI aids.
This work opens the door to designing behavior-based AI aids and better understanding how humans represent, develop, and update mental models of AI systems.

\begin{acks}
This material is based upon work supported by an Amazon grant, a National Science Foundation grant under No. IIS-2040942, and the National Science Foundation Graduate Research Fellowship Program under grant No. DGE-1745016. Any opinions, findings, and conclusions or recommendations expressed in this material are those of the authors and do not necessarily reflect the views of Amazon or the National Science Foundation.
\end{acks}

\bibliographystyle{ACM-Reference-Format}
\bibliography{references-zot}

\appendix
\section{Additional Study Details}

\begin{figure*}
    \centering
    \includegraphics[width=\textwidth]{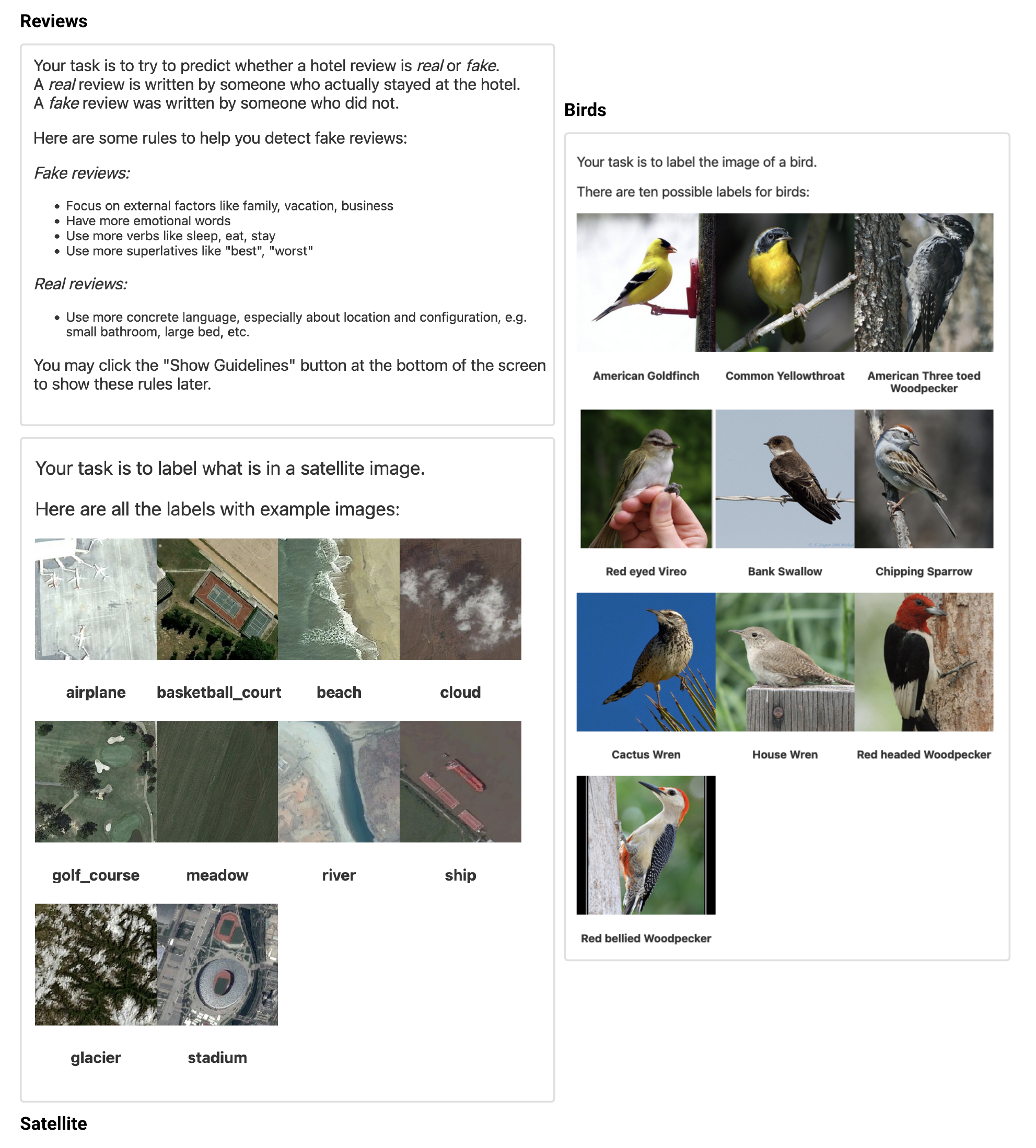}
    \caption{
    The instructions shown to participants for each task before labeling.
    }
    \label{fig:instructions}
\end{figure*}

To improve the reproducibility of this work, we include additional details about the study here.
Table \ref{tab:parts} shows the final count of participants per condition after removing those who did not pass the attention check.
Table \ref{tab:bds} shows the specific language used for each behavior description in each task.
Lastly, Figure \ref{fig:instructions} shows the specific instructions shown to participants before labeling for each task and the subsets of labels used in the two multiclass tasks.


\begin{table}
\caption{The number of participants per condition after removing failed attention checks.}
\begin{tabular}{|l |c c c|}
 \hline
  & Reviews & Satellite & Birds \\
 \hline
 No AI & 24 & 23 & 22 \\ 
 \hline
 AI & 23 & 24 & 23 \\
 \hline
 AI + BD & 23 & 25 & 25 \\
 \hline
\end{tabular}
\label{tab:parts}
\end{table}

\begin{table}
\caption{The specific behavior descriptions used for each behavior description group.}
\begin{tabular}{|p{0.09\textwidth} | p{0.25\textwidth}| p{0.25\textwidth}| p{0.25\textwidth}|}
 \hline
  & Reviews & Satellite & Birds \\
 \hline
 Group 1 & The AI is only 40\% accurate for reviews like this with less than 50 words. & The AI confuses golf courses and meadows 60\% of the time. & The AI confuses Cactus Wrens with House Wrens 60\% of the time. \\ 
 \hline
 Group 2 & The AI is only 20\% accurate for reviews like this with more than 3 exclamation marks. & The AI confuses glaciers and clouds 80\% of the time. & The AI confuses Read Headed Woodpecker with Red Bellied Woodpecker 80\% of the time. \\
 \hline
\end{tabular}
\label{tab:bds}
\end{table}

\end{document}